\documentclass[
  journal=pasa,
  manuscript=research-article, 
  year=2025,
  volume=42,
]{cup-journal}

\usepackage{amsmath}
\usepackage[nopatch]{microtype}
\usepackage{booktabs}

\usepackage{aas_macros}
\usepackage{hyperref}
\usepackage{siunitx}
\usepackage{svg}
\usepackage{graphicx}
\usepackage{multicol}
\usepackage{subcaption}
\usepackage{overpic}
\DeclareSIUnit{\solidradian}{sr}
\usepackage{soul}
\usepackage[export]{adjustbox}
\usepackage[most]{tcolorbox}

\hypersetup{colorlinks,citecolor=blue,linkcolor=blue,urlcolor=blue}

\received {25 June 2025}
\revised  {22 August 2025}
\accepted {04 September 2025}
\published{03 October 2025}

\newcommand{\var}[2]{\newcommand{#1}{#2}}
\var{\figw}{1.0}

\newcommand{\fig}[4][1.0] 
{
    \begin{figure}[h]
        \centering
        \includegraphics[width=#1\linewidth]{#3}
        \caption{#4}
        \label{fig:#2}
    \end{figure}
}

\newcommand{\eqn}[2] 
{
    \begin{equation}
        \label{eqn:#1}
        \begin{aligned}
            #2
        \end{aligned}
    \end{equation}
}

\usepackage{booktabs}
\usepackage{array}
\newcolumntype{R}{>{\raggedleft\let\newline\\\arraybackslash}}
\newcolumntype{L}{>{\raggedright\let\newline\\\arraybackslash}}

\newcommand{\tab}[5][H] 
{
    \bgroup
        
        \setlength\tabcolsep{10pt}
        \begin{table}[#1]
            \begin{center}
                \begin{tabular}{#3}
                        #4
                \end{tabular}
            \end{center}
            \caption{#5}
            \label{tab:#2}
        \end{table}
    \egroup
}

\usepackage{siunitx}
\DeclareSIUnit{\solidradian}{sr}
\DeclareSIUnit{\log}{ln}
\DeclareSIUnit{\jy}{Jy}

\newcommand{\DMunit}{pc\,cm\ensuremath{^{-3}}}
\newcommand{\farcs}{\mbox{$.\!\!^{\prime\prime}$}}

\newcommand{\PSRpos}{PSR\,J0410\ensuremath{-}31}
\newcommand{\GLXLPT}{GLEAM-X\,J0704\ensuremath{-}37}
\newcommand{\psrB}{PSR\,J0031\ensuremath{-}57}
\newcommand{\psrF}{PSR\,J0034\ensuremath{-}0721}
\newcommand{\psrE}{PSR\,J2048\ensuremath{-}1616}
\newcommand{\psrC}{PSR\,J0437\ensuremath{-}4715}
\newcommand{\psrA}{PSR\,J0630\ensuremath{-}2834}
\newcommand{\psrG}{PSR\,J2241\ensuremath{-}5236}
\newcommand{\psrH}{PSR\,J0502\ensuremath{-}6617}
\newcommand{\psrI}{PSR\,J1244\ensuremath{-}1812}

\title{A long period transient search method for the Murchison Widefield Array}

\author{Csan\'{a}d Horv\'{a}th}
\affiliation{International Centre for Radio Astronomy Research, Curtin University, Kent St, Bentley WA 6102, Australia}
\email[Csan\'{a}d Horv\'{a}th]{csanad.horvath@postgrad.curtin.edu.au}

\author{Natasha Hurley-Walker}
\affiliation{International Centre for Radio Astronomy Research, Curtin University, Kent St, Bentley WA 6102, Australia}

\author{Samuel J. McSweeney}
\affiliation{International Centre for Radio Astronomy Research, Curtin University, Kent St, Bentley WA 6102, Australia}

\author{Timothy J. Galvin}
\affiliation{Commonwealth Scientific and Industrial Research Organisation (CSIRO), 26 Dick Perry Avenue Kensington WA 6151, Australia}

\author{John Morgan}
\affiliation{Commonwealth Scientific and Industrial Research Organisation (CSIRO), 26 Dick Perry Avenue Kensington WA 6151, Australia}


\doi{\href{https://doi.org/10.1017/pasa.2025.10093}{10.1017/pasa.2025.10093}}

\begin{document}

\begin{abstract}
We present an automated search method for radio transients on the minute timescale focused on the emerging long period transients (LPTs) in image-plane radio data. The method is tuned for use with the Murchison Widefield Array (MWA) and tested on archival observations from the GaLactic and Extragalactic All-Sky MWA Extended Survey (GLEAM-X) in the 70--300\,MHz range. The images are formed from model-subtracted visibilities, before applying three filters to the time series of each pixel in an image, with each filter designed to be sensitive to a different transient behaviour. Due to the nature of radio interferometry and the refraction of the fluctuating ionosphere, the vast majority of candidates at this stage are artefacts which we identify and remove using a set of flagging measures. Of the 336 final candidates, 7 were genuine transients; 1 new LPT, 1 new pulsar, and 5 known pulsars. The performance of the method is analysed by injecting modelled transient pulses into a subset of the observations and applying the method to the result.
\end{abstract}


\section{Introduction}\label{sec:introduction}

Transient astrophysical events typically require highly energetic environments, providing opportunities to study otherwise inaccessible physical regimes.
Some transient phenomena are unique to radio or obscured by dust at other frequencies, and many radio transients feature emission from the most extreme environments in the universe such as pulsars, magnetars, rotating radio transients, fast radio bursts, gamma ray bursts, active Galactic nuclei, and long period transients (LPT).

LPTs are a recently discovered class of object which produce beamed radio emission characterised by their extremely long minute to hour timescale periods, complex temporal structure, and wide 10-s to 20-min pulse widths \citep{dong2024discovery, lee2025emission}.
Their long periods combined with their slow spin-down rate make them difficult to reconcile with existing models as there is not enough angular momentum to power their emission \citep{rea2024radio}.
There are only a handful of discovered LPTs, some of which are confirmed to be white dwarf (WD) binary systems \citep{rodrigues2025, de2025sporadic} while others \citep{Hurleywalker2022, Hurleywalker2023, dong2024discovery, lee2025emission, wang2025detection} have no confirmed counterpart at other frequencies and could be WD binaries, neutron stars, or other exotic objects.

LPTs represent a new parameter space for transient searches at radio frequencies, requiring tailored survey specifications and data processing methods.
Looking for objects which appear for $\sim$\qty{30}{\second} every $\sim$hour and often cease emission for extended periods of time requires long dwell times, well-filled snapshot $(u,v)$-coverage, and a wide field-of-view (FOV) to have a chance at surveying an appreciable portion of the sky. At these timescales only single-pulse searches are practicable, and so far all LPTs have been discovered through imaging searches by interferometers. It is difficult to detect such sources using single-dish telescopes due to the presence of red noise on this timescale, which is typically subtracted by gain control systems.
The Murchison Widefield Array (MWA) -- used in this work -- is a radio telescope well suited to this task operating in the 70--300\,MHz frequency range with a 30.72\,MHz processed bandwidth. It has a field of view of $\sim$300 to $\sim$1600 square degrees in the 215\,MHz and 87\,MHz frequency bands respectively \citep{tingay2013murchison}.

Two recent searches for radio transients and variables similar to this work are described by \citet{de2024transient} and \citet{yuanming2023} using the LOFAR Two-metre Sky Survey (LoTSS) at 144\,MHz and the Australian SKA Pathfinder (ASKAP) pilot surveys at 1 GHz respectively.
In both, model visibilities were subtracted from the observations with the goal of subtracting the non-variable components, and the result was imaged.
\citeauthor{de2024transient} generated 8-s snapshot images from their model-subtracted observations and performed source-finding on the results.
\citeauthor{yuanming2023} generated 15-min snapshot images from their 8 to 10 hour model-subtracted observations. From the image cubes, they generated three statistical maps for each observation: chi-squared, peak value, and Gaussian correlation, to which they applied source-finding.
Visibility differencing was also used by \citet{dobie2022}, \citet{caleb2024emission}, and \citet{fijma2024} to similar effect.
The resulting images contain the variable and transient components of the continuum, but they are also contaminated by scintillating non-variable sources, synthesised beam sidelobes, aliases, and other imaging artefacts.

Ionospheric scintillation is especially challenging due to the overwhelming quantity of non-variable sources, and the fact that the fluctuating ionosphere of the Earth varies the apparent position of those sources on the sky \citep{Waszewski_Morgan_Jordan_2022}.
To combat this, \citet{de2024transient} defined a list of mask radii for ranges of artefact flux densities around known sources.
Similarly, \citeauthor{yuanming2023} rejected all candidates within 30 arcseconds of a known source unless they varied significantly when compared with the known brightness of the associated source.
Filtering out artefact candidates is a cornerstone of this work because it is crucial to improving the sensitivity of radio-transient surveys without being overwhelmed by false-positives. 

In this work we present a method to search for radio transients on the second to minute timescales in MWA data. Here, we test the method on 7099 archival observations from GLEAM-X data releases I and II (DRI and DRII) \citep{2022PASA...39...35H,2024PASA...41...54R}, although results of the method \citep{Hurleywalker2023,2025MNRAS.tmp.1156M} from other Galactic plane MWA data (Hurley-Walker et al. in prep.) already exist. The method design is focused on the properties of known LPTs (although limited by the observation lengths), but other transient and variable sources are also detected. These observations are not concentrated on the Galactic plane where the highest number density of LPTs is expected, but they served as an accessible and extensive dataset for development purposes. Despite this, we found a new LPT \GLXLPT{} (\autoref{subsec:gleamxj0704-37}) \citep{Hurleywalker2024} and a new pulsar \psrB{} (\autoref{subsec:psrj0031-37}) \citep{Mcsweeney_2025}.
The transient search pipeline is described in \autoref{sec:method}, composed of model-subtraction, artefact flagging, self-cross-matching, and visual inspection. Resulting transient candidates are presented in \autoref{sec:results}, and the effectiveness of the search method is discussed in \autoref{sec:evaluation_of_performance}.

\section{Data}\label{sec:data}

The data used in this work, GLEAM-X Data Releases I and II, described fully by \citet{2022PASA...39...35H} and \citet{2024PASA...41...54R} respectively, is a low frequency continuum survey by the MWA covering the sky south of Dec $+30$, broken into seven drift scans. The frequency coverage is composed of five bands between 72 and 231\,MHz, avoiding the Orbcomm satellite band by excluding the band around 137\,MHz. The observing was executed as a series of week-long campaigns between 2018 and 2020, where each night produced 120-s scans for each of the five frequency settings for a single declination.
The GLEAM source catalogue \citep{Hurleywalker2016} was used in the cross-matching step (\autoref{subsec:crossmatch}) because it is sensitive enough to find all the sources in the 4-s integrations of GLEAM-X (\autoref{subsec:imaging}).

\section{Search Method}\label{sec:method}

The transient search pipeline looks for transient candidates by applying filters to image cubes. 
The image cubes consist of two spatial dimensions and one temporal dimension. The sky model is subtracted from the cube to remove non-variable sources as described in \autoref{subsec:imaging}. Three filters, described in \autoref{subsec:the_filters}, are applied to each temporal column to generate 2D filter images. They are intended to probe different timescales and pulse morphologies. Candidate transients are found by looking for islands of large values in the filter maps, described in \autoref{subsec:island_detection}. Candidates are cross-matched with the GLEAM catalogue in \autoref{subsec:crossmatch} and those likely to be artefacts rather than genuine transients are flagged according to the criteria in \autoref{subsec:candidate_selection}. The catalogue of remaining un-flagged candidates is cross-matched with itself, and a S/N cutoff is applied which decreases for candidates with multiple detections in \autoref{subsec:grouping}. The pipeline is written in Python and makes use of the Astropy library \footnote{\url{https://www.astropy.org/}} and parts of the GLEAM-X imaging pipeline.

\subsection{Imaging}\label{subsec:imaging}

\begin{figure}[hbt!]
    \centering
    \includegraphics[width=\linewidth]{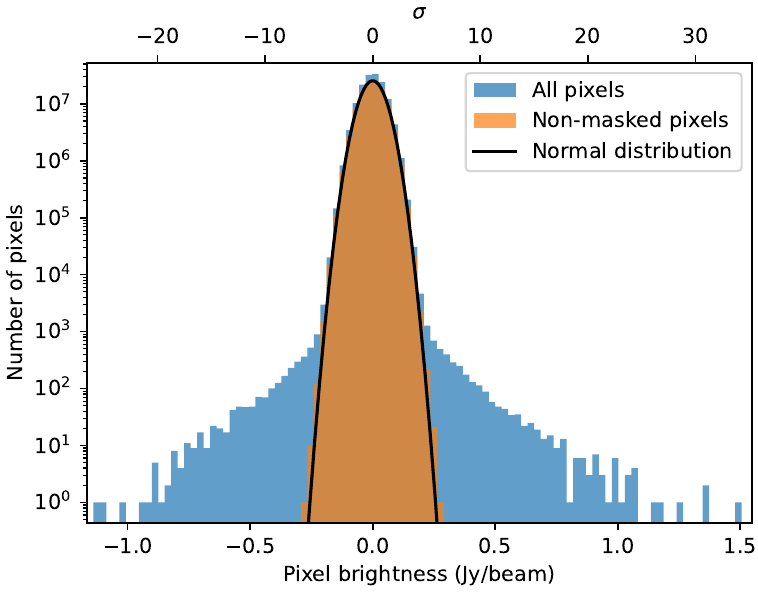}
    \caption{Pixel brightness distribution of a representative 154\,MHz model-subtracted data cube, as well as the same data with a 4$'$ mask applied around the sources in the GLEAM catalogue. The thermal noise dominating the data is approximately normally distributed with $\sigma = \text{RMS}$. The deviation from normal above $\sim5\sigma$ is primarily due ionospheric scintillation of real radio sources.}
    \label{fig:noise}
\end{figure}

The formation of the image data is described by \citet{2022PASA...39...35H}; to summarise, deep images and visibility models are formed by deconvolution of each two-minute snapshot; these model visibilities are then subtracted and every 4-s interval of a given snapshot observation is imaged. The data are stored as HDF5 cubes which are largely thermal-noise-dominated as shown in \autoref{fig:noise}, storing only the differences between each time step and the continuum average.
Note that the cubes are not cleaned or primary beam corrected.

Spatially, the cubes are $2400 \times 2400$\,pixels covering 1742\,deg$^2$ to 290\,deg$^2$ with point spread function (PSF) full width at half maximum (FWHM) semi-major axes of $156''$ to $62''$ for the 88\,MHz to 216\,MHz bands respectively. The PSF is slightly under-sampled with 2.5\,pixels across the semi-major axis to reduce computational cost. For more details see \autoref{tab:data_properties}.
These cubes form the basis of the searches described in this paper.

RFI sometimes overwhelms a few time-steps in an observation, so time-steps whose root-mean-squared (RMS) was greater than 1.5 $\times$ the cube RMS were dropped before applying the transient filters.

\subsection{Transient Filters}\label{subsec:the_filters}

The following transient filters take as input a model-subtracted observation cube and output a 2D image whose values aim to roughly scale with how variable the brightness of a pixel is.

    \subsubsection{Spike}
    The Spike filter is intended to look for unresolved bursts narrower than the bin width of \qty{4}{\second}. A sigma-clip $\sigma_\text{clip} = 4$ is applied to the time-series of each pixel to estimate the noise level (\autoref{eqn:spike1}), and the maximum number of standard deviations above the clipped pixel time-series is recorded as the Spike value at that pixel (\autoref{eqn:spike2}).
    \begin{equation}\label{eqn:spike1}
        I_\text{noise} = \left\{J \in I ~\middle|~ \frac{J-\bar{I}}{\text{std}(I)} \leq \sigma_\text{clip}\right\}
    \end{equation}
    \begin{equation}\label{eqn:spike2}
        \text{spike}(I) = \text{max} \left( \frac{I-\bar{I}_\text{noise}}{\text{std}(I_\text{noise})} \right)
    \end{equation}
    where $I$ is the time series of the pixels at a coordinate.
    
    \subsubsection{Time Correlated Gaussian (TCG)}
    The TCG filter looks for wider pulses on the order of 10-s. It is a matched-filter that correlates the time-series of a cube at each pixel with a Gaussian kernel with standard deviation $\sigma_\text{tcg}$ to find approximately Gaussian shaped pulses.
    \begin{equation}\label{eqn:tcg1}\begin{aligned}
        K_i = \left\{ \begin{array}{cl}
        \text{exp}\left(-\frac{t_i^2}{2\sigma_\text{tcg}^2}\right) & \text{for} ~ -\frac{W}{2} < t_i < \frac{W}{2} \\
        0 & \text{elsewhere}
        \end{array} \right.
    \end{aligned}\end{equation}
    \begin{equation}\label{eqn:tcg2}
        \text{tcg}(I) = \left(\frac{K-\bar{K}}{\sum_{i=1}^n\left(K_i-\bar{K}\right)^2}\right) * \left(I-\bar{I}\right)
    \end{equation}
    where $K=\lbrace K_1,...,K_n\rbrace$ is the kernel, $t$ is time in seconds since the start of the observation, $W$ is the width of the kernel, $i$ is the index of the time-step, and $n$ is the number of time-steps. The $*$ represents the convolution operation.
    The choice of parameter values are shown in \autoref{tab:filter_settings} and discussed in \autoref{sec:evaluation_of_performance}.
    
    \subsubsection{Root-mean-square (RMS)}
    The RMS filter simply calculates the RMS through time of each pixel as
    \begin{equation}\label{eqn:rms}
        \text{rms}(I) = \sqrt{\frac{1}{n} \sum_i \left( I_i - \bar{I} \right)^\text{2}}
    \end{equation}
    It is intended to detect highly variable, stochastic sources without a clear pulse shape. The island thresholds for the RMS filter are proportional to the cube RMS as per \autoref{tab:filter_settings}.

\subsection{Island Detection}\label{subsec:island_detection}

A 2D boolean map is generated which is true where any of the three filters are above the lower threshold (\autoref{tab:filter_settings}). Islands are then found using a 3 by 3 connectivity matrix of all ones.
The choice of lower threshold is a balance between flood-filling the entire image and having enough pixels per island for a defined shape.

Islands whose maximum TCG, Spike, and/or RMS filter values are greater than the upper threshold are assigned the valid\_tcg, valid\_spike, and/or valid\_rms flags respectively. Islands which do not have either of the aforementioned flags are immediately discarded.
The upper threshold is a balance between the true and false positive rates, but it was chosen purposefully low such that the bottom candidates are entirely unconvincing so that stricter selection criteria can be chosen in later steps without redoing the island detections. Each island is fitted with the smallest area ellipse which encloses all of the island pixels, defined by the ellipse center, major and minor radii, and rotation angle.

\begin{table}[hbt!]
    \caption{
        Transient search filter settings. The lower threshold is the threshold for island area flood-filling. The upper threshold is the minimum peak pixel brightness for a candidate to be considered. ``Cube RMS'' refers to the RMS of the transient cube. $A = \left\lbrace 1.5 \text{ for 87\,MHz; } 1.25 \text{ for 118\,MHz; } 1 \text{ otherwise.} \right\rbrace$
    }
    \label{tab:filter_settings}
    \begin{tabular}{llll}
        \toprule
        \headrow Filter name & Lower threshold       & Upper threshold       & Parameters               \\
        \midrule
                 TCG         & 5.5 $\times$ Cube RMS & 7   $\times$ A $\times$ Cube RMS & $\sigma_\text{tcg} = 4s$ \\
                             &                       &                                  & $W = 100s$    \\
                 Spike       & 5.5                   & 8.5                   & $\sigma_\text{clip} = 4$ \\
                 RMS         & 2 $\times$ Cube RMS   & 2.5 $\times$ A $\times$ Cube RMS &                          \\
        \bottomrule
    \end{tabular}
\end{table}

Several artefacts which would ordinarily be removed by standard imaging techniques are still present in our image cubes because the 4-s intervals have not been individually cleaned due to the computational expense (see \autoref{subsec:imaging}). We therefore have to ensure that the selection criteria described in subsequent sections are able to identify the following broad classes of artefacts which dominate the candidates at this stage:
\begin{enumerate}
    \item Ionospheric scintillation: The ionosphere is constantly in motion with waves of electron density on the scale of tens of metres to tens of kilometres. The varying electron density both diffracts and refracts radio waves causing amplitude and phase scintillation which increases at longer wavelengths. The neighbourhoods of non-variable sources are frequently falsely detected as their apparent positions shift in and out of neighbouring pixels \citep{jordan2017characterization, hurley2018distorting}. Additionally, amplitude scintillation from both the ionosphere and the interplanetary medium can impose flux variability on otherwise non-variable sources \citep{Waszewski_Morgan_Jordan_2022}.
    
    \item Synthesised Beam Sidelobes:
    In radio imaging, the sky is convolved with the synthesised beam which, if the calibration is imperfect, results in subtracted sources being surrounded by extended artefacts. Due to scintillation of the parent source and the rotation of the sky relative to the primary beam, the variation of the sidelobe artefacts (which exhibit structure on spatial scales comparable to the PSF), can be erroneously detected as transients.
    
    \item Aliases: The primary beam of the MWA tiles has sidelobes with sensitivity $\sim10\%$ the main beam response \citep{Sokolowski_2017}, so bright ($>$100-Jy) sources outside of the field of view can make significant contributions to the visibilities \citep[e.g.][section 3.2]{1999ASPC..180..127B}; such sources are aliased into the image, and often appear to move. It is prohibitively expensive to image the whole sky, but the peeling algorithm \citep{Hurleywalker2016} removes most aliases by subtracting a model of bright sources from the visibilities. Assuming an accurate sky model, this method is effective, but the computational expense prohibits the perfect peeling of all sources. Notwithstanding efforts to suppress aliasing, sources from outside the field of view can be detected.
    
    \item Radio frequency interference (RFI): Interference from terrestrial sources of radiation such as satellites and aeroplanes interfere with the antennas, usually in short bursts. Generally, RFI is easily identified in the image plane as diagonal stripes as one baseline, and therefore spatial frequency, overwhelms the rest.

    \item Thermal noise: The background of the images is full of approximately Gaussian thermal noise (see \autoref{fig:noise}). Along the time axis, MWA noise is remarkably clean of systematic effects. The thermal noise is not correlated between time-steps, tending to result in sparse low (S/N) Spike candidates. 
\end{enumerate}

\subsection{Catalogue Cross-Matching}\label{subsec:crossmatch}

In the interest of flagging scintillating known sources, the catalogue of islands is cross-matched with the GLEAM catalogue. It is sometimes ambiguous which source an island is associated with in its neighbourhood, so the following two possible matches are recorded for each candidate:
the nearest source to the candidate, and the brightest source within the greater of 4$'$ (the maximum scale at which we observe position shifts due to ionospheric scintillation) and the island's radius.
Throughout the rest of the paper, ``nearest known source'' refers to either of the aforementioned, and conditions involving the ``nearest known source'' are applied to both.

\subsection{Flagging of Artefacts}\label{subsec:candidate_selection}

Candidates are assigned flags by the below criteria. Candidates for which any of the following flags are true are rejected. The decision making process which lead to the design of the flags and their specific parameters is discussed in \autoref{sec:evaluation_of_performance}.

\begin{figure}[hbt!]
    \centering
    \includegraphics[width=\linewidth]{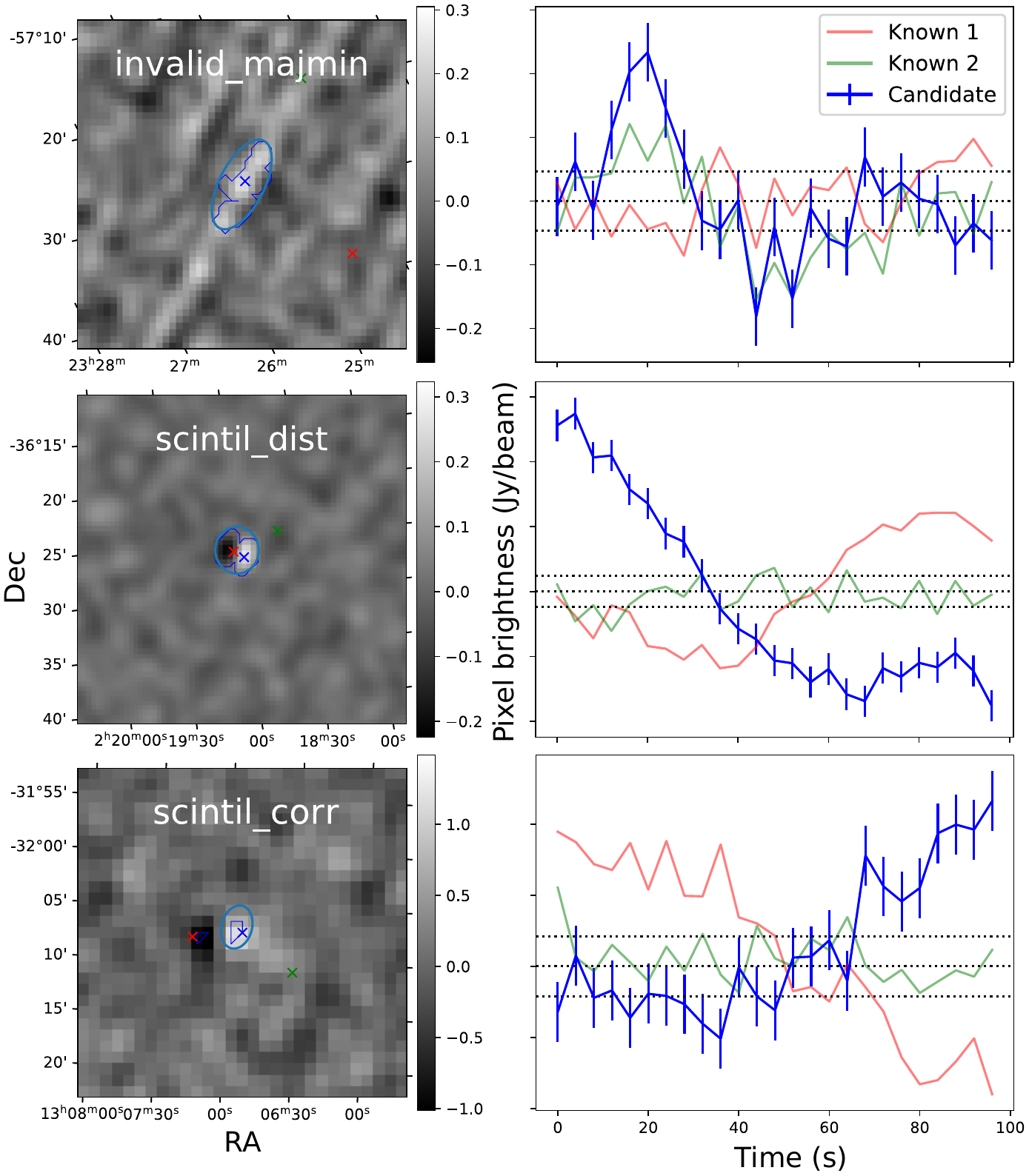}
    \caption{Examples of candidates which were excluded due to the invalid\_majmin, scintil\_dist, or scintil\_corr flags alone. At left are 0.5\textdegree{} cutouts of the time-step of the transient cubes corresponding with the maximum brightness of the candidates. The blue cross marks the coordinates of the peak, the red and green crosses mark the two nearest known sources. At right are the lightcurves at the marked coordinates. The dark blue contour marks the flood-filled island, to which the light blue ellipse is fitted. The horizontal dotted lines mark the mean and RMS of the transient cube.}
    \label{fig:example_cutouts}
\end{figure}

\begin{enumerate}
    \item \verb|invalid_beam|: Is the primary beam at the candidate coordinates less than 50\% of the maximum value within the observation? The edges of observations have too low sensitivity to be useful.
    \item \verb|invalid_majmin|: Is the ratio between the major and minor radii of an ellipse fitted to contain the island greater than 2, AND does the island include more than 3 pixels? This flag is intended to catch extended imaging artefacts which appear either radially around bright stationary sources, or as a result of a moving source. Moving sources can be aliases from outside the field of view, or real sources like airplanes or satellites \citep{detection2013tingay}. For an example of a candidate with this flag, see \autoref{fig:example_cutouts}. Limiting this flag to sources with at least 3 pixels prevents the rejection of candidates with islands too small to have a meaningful shape.
    \item \verb|invalid_area|: Does the candidate island have more than 20 pixels? Aliases often appear as distorted patches rather than point sources.
    \item \verb|invalid_mean|: Is the mean brightness of the candidate greater than the RMS of the observation? Sources or aliases which are not completely removed in the model subtraction step tend to have flat, positive lightcurves.
    \item \verb|scintil_dist|: Is the separation between the candidate and any of the nearest known sources less than the greater of 4$'$ and the major island radius, AND
    \begin{itemize}
        \item is it an 87\,MHz observation AND is the candidate peak beam-corrected flux density less than 1.5 $\times$ the nearest known source's flux density, OR
        \item is it a 118\,MHz observation AND is the candidate peak beam-corrected flux density less than 1.25 $\times$ the nearest known source's flux density, OR
        \item is the candidate peak beam-corrected flux density less than the nearest known source's flux density?
    \end{itemize}
    Here, we assume that candidates are unresolved, subtending at most one synthesised beam, and thus implicitly convert from pixel brightness (Jy/beam) to flux density (Jy). We then correct for the primary beam to allow comparison with the catalogue.
    
    In the low frequency regime, ionospheric scintillation often boosts the apparent brightness of sources. A 4$'$ circular mask was deemed as a good compromise between loss of sky area and reduction of false positives due to scintillation of non-variable sources. Candidates are not masked if they are brighter than their masking known source, as that is unlikely to be scintillation. For an example of a candidate with this flag, see \autoref{fig:example_cutouts}.
    \item \verb|scintil_corr|: Is the separation between the candidate and any of the nearest known sources less than the greater of 8$'$ and twice the major island radius, AND is the known source brighter than the candidate peak beam-corrected brightness, AND is the lightcurve at the known source coordinate negatively correlated with the candidate lightcurve (Pearson coefficient > 0.85)? If ionospheric scintillation shifts the apparent position of a source by more than 4$'$, then the known source is usually in the negative lobe of the artefact (red in row 3 of \autoref{fig:example_cutouts}), which is negatively correlated with the positive lobe (blue in row 3 of \autoref{fig:example_cutouts}).
    \item \verb|close_to_ateam|: Is the candidate within 5\textdegree{} of any of the following sources? These bright sources produce imaging artefacts too severe to analyse in a more careful way than a simple circular mask.
    \begin{itemize}
        \item \makebox[3cm]{Centaurus A \hfill} (13:25:28, $-$43:01:09)
        \item \makebox[3cm]{Hydra A     \hfill} (09:18:06, $-$12:05:44)
        \item \makebox[3cm]{Pictor A    \hfill} (05:19:50, $-$45:46:44)
        \item \makebox[3cm]{Hercules A  \hfill} (16:51:08, $+$04:59:33)
        \item \makebox[3cm]{Virgo A     \hfill} (12:30:49, $+$12:23:28)
        \item \makebox[3cm]{Crab        \hfill} (05:34:32, $+$22:00:52)
        \item \makebox[3cm]{Cygnus A    \hfill} (19:59:28, $+$40:44:02)
        \item \makebox[3cm]{Cassiopeia A\hfill} (23:23:28, $+$58:48:42)
    \end{itemize}
    \item \verb|close_to_bright|: Is the candidate within 1\textdegree{} of a source brighter than 10 Jy in the GLEAM source catalogue?
    \item \verb|is_moon|: Is the candidate within 4$'$ of the moon? The moon reflects Earth's RFI, which is sometimes detected as a transient candidate, particularly in the 118\,MHz and 154\,MHz observing bands \citep{mckinley2013_moon_rfi}.
    \item \verb|invalid_freq|: Is the candidate from an observation centred on 87\,MHz? We decided to exclude the bottom frequency band due to the prevalence of extreme ionospheric scintillation making it difficult to draw useful conclusions. Further justification is given in \autoref{sec:evaluation_of_performance}.
    \vspace{1px}
\end{enumerate}

\subsection{Candidate Grouping}\label{subsec:grouping}

The goal of the preceding section is not to build a perfect classifier, but rather to minimise the number of candidates that need to be visually inspected while also minimising the risk of discarding a candidate which would be interesting enough to warrant further investigation if it were to be visually inspected.
At this stage, we have eliminated the most obvious false positives, but we have not yet incorporated mutli-observation information into the automated part of the decision making.
A low-S/N candidate which only appears once is unlikely to be real, but if an equally dim candidate appears multiple times then it may be worth investigating. In practice, this means that the S/N requirements for repeating candidates can be lowered. 

The candidate database was cross-matched with itself with a cross-match radius of 4$'$ to create candidate `groups'. For each group, the mean peak pixel brightness S/N and the mean pixel fluence S/N was calculated. Fluence was chosen in addition to the peak pixel brightness to bias pulses which may not be so bright but have a longer duration (and so are equally as convincing as a shorter bright pulse).

We calculate the peak pixel brightness S/N $P$ as
\begin{equation}
    P = \frac{I_k}{\sigma_I} = \frac{N_k + R_k}{\sigma_I}
\end{equation}
where $N$ and $R$ are the noise and real contributions to the pixel brightness respectively, and $k$ is the time-step index at which the pixel brightness is at maximum. We assume $N$ is normally distributed about 0 with standard deviation $\sigma_I \simeq \text{cube RMS}$. The mean peak pixel brightness S/N of a group of $m$ candidates is
\begin{equation}
    \bar{P} = \frac{1}{m}\sum_{j=1}^m P_j = \frac{1}{m}\sum_{j=1}^m \left( (N_k)_j + (R_k)_j \right) = \bar{P}_N + \bar{P}_R
\end{equation}
where $j$ is the candidate index in the group, and $\bar{P}_N$ and $\bar{P}_R$ are the noise and real contributions to $\bar{P}$. The standard deviation of $\bar{P}_N$ across many groups of size $m$ will be $\sigma_{\bar{P}} = \frac{1}{\sqrt{m}}$ because $\sigma_P = 1$ by definition. We choose a cutoff $C_P$ and select candidate groups whose $\bar{P}$ is more than $C_P$ times the noise level according to
\begin{equation}
    \bar{P} > C_P \sigma_{\bar{P}} = \frac{C_P}{\sqrt{m}}
\end{equation}
For a group size of 1 this reduces to a simple S/N cutoff.

The cutoff for the group mean pixel fluence S/N is derived similarly. We calculate the pixel fluence $F$ as
\begin{equation}
    F = \Delta t\sum_{i=1}^{n}I_i = \Delta t\sum_{i=1}^{n}(N_i + R_i) = F_N + F_R
\end{equation}
where $n$ is the number of time-steps in the observation, $F_N$ and $F_R$ are the noise and real contributions to the pixel fluence respectively, and $\Delta t$ is the time step. $F_N$ will have a standard deviation across many sources of $\sigma_F = \sigma_I \Delta t \sqrt{n}$. Hence, we define the pixel fluence S/N $f$ as
\begin{equation}
    f = \frac{F}{\sigma_I \Delta t \sqrt{n}}
\end{equation}
and the group mean pixel fluence S/N as
\begin{equation}
    \bar{f} = \frac{1}{m}\sum_{j=1}^m f_j = \frac{1}{m}\sum_{j=1}^m (f_{Nj} + f_{Rj}) = \bar{f}_N + \bar{f}_R
\end{equation}
where $f_N$ and $f_R$ are the noise and real contributions to $f$ respectively. The standard deviation of $\bar{f}_N$ across many groups of size $m$ of will be $\sigma_{\bar{f}} = \frac{1}{\sqrt{m}}$ because $\sigma_f = 1$ by definition. We choose a cutoff $C_F$ and select candidate groups whose $\bar{f}$ is more than $C_F$ times the noise level according to
\begin{equation}
    \bar{f} > C_F \sigma_{\bar{f}} = \frac{C_F}{\sqrt{m}}
\end{equation}

This is a simplistic approach, but for the purpose of creating a reasonable cutoff curve for candidate believability, it is adequate. 
A value of $C_P = C_F = 7$ was chosen, and candidates which fall below both cutoffs are rejected. The cutoff curves are plotted in \autoref{fig:group_size_vs_group_sn}. The value of $7$ aligns the most closely with the visual believability of candidates, and can be interpreted as a more advanced $7\sigma$ signal requirement. 

\subsection{Classification Interface}\label{subsec:interface}

All candidates which make it through the above cuts need to be visually inspected to determine suitability for further investigation.
A candidate classification web-interface was developed by Astronomy Data and Computing Services (ADACS).
The interface presents the user with a diagnostic plot, such as the one for \GLXLPT{} in \autoref{fig:glxlpt}, as well as a GIF cutout, cutouts from the three filter maps, significance of the candidate's brightness relative to the rest of the observation, further diagnostic information about the candidate and the observation, options to categorise and rate the candidate, other nearby candidates, and nearby objects from the SIMBAD and ATNF catalogues. This was an invaluable resource both to generate the final shortlist of candidates, and during development to tune the parameters.

\section{Results}\label{sec:results}

\begin{figure}[hbt!]
    \centering
    \includegraphics[width=\linewidth]{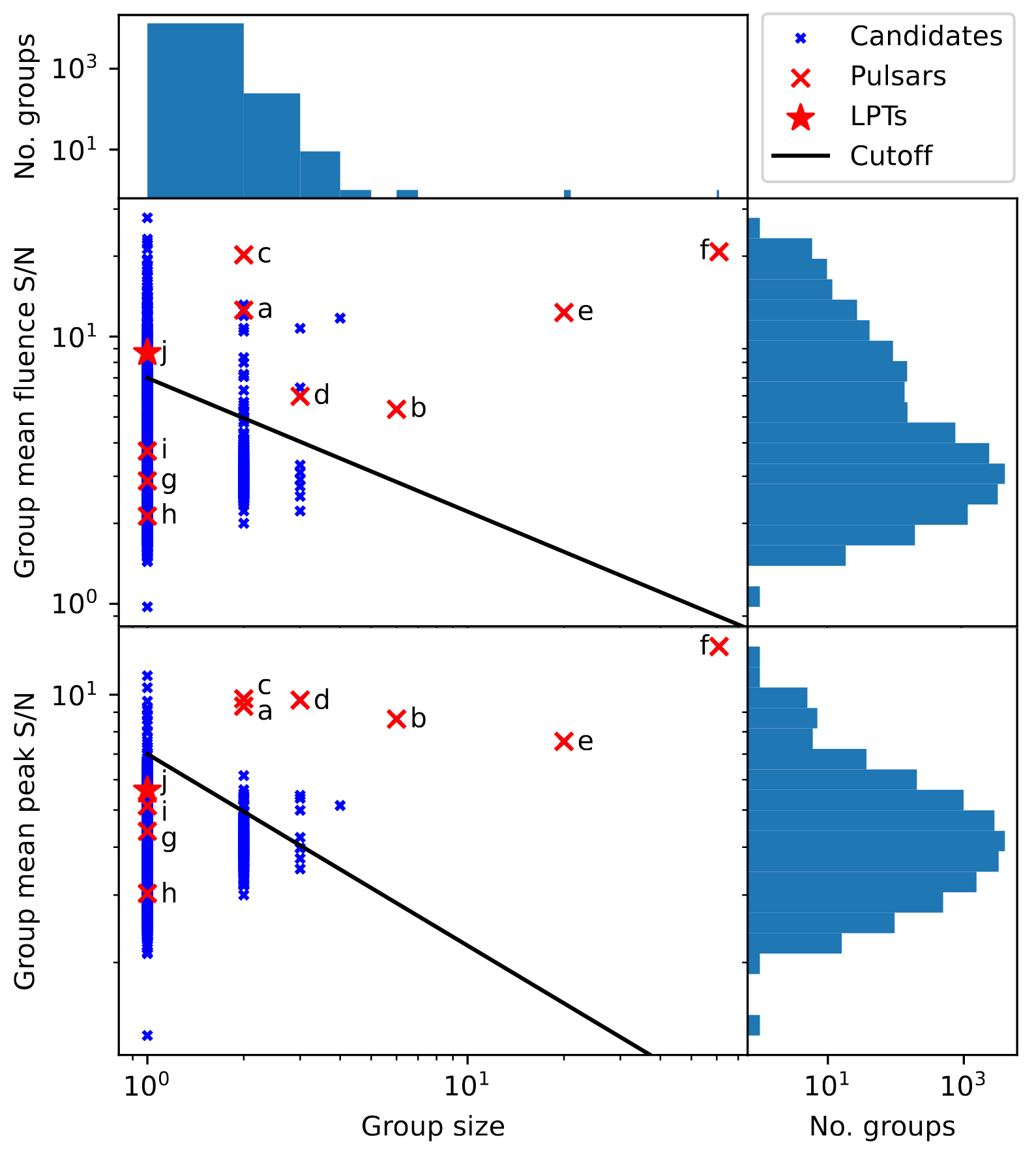}
    \caption{Distribution of mean pixel fluence and peak pixel brightness S/N statistics against group size. The cutoff curve is drawn on both. The groups marked in red are identified as the following real sources: a. \psrA{}; b. \psrB{}; c. \psrC{}; d. \PSRpos{}; e. \psrE{}; f. \psrF{}; g. \psrG{}; h. \psrH{}; i. \psrI{}; j. \GLXLPT{}.}
    \label{fig:group_size_vs_group_sn}
\end{figure}

Of the 13844 candidates found in the dataset, 610 formed groups of 2 or more, making 13493 total groups. Of those, there are 330 candidate groups above the fluence S/N cutoff in \autoref{fig:group_size_vs_group_sn}, and 33 above the peak pixel brightness S/N cutoff. Only 6 of the peak pixel brightness S/N groups are not among the 330 pixel fluence S/N groups, and none of them passed visual inspection. This makes the final number of candidate groups 336, 7 of which were found to be genuine transients. One source was the real LPT \GLXLPT{} (see \autoref{subsec:gleamxj0704-37}), detected by TCG. The following 6 were found to be pulsars:
\begin{enumerate}[label=\alph*.]
    \item \makebox[3.5cm]{\psrA{}   \hfill} (TCG \& RMS)         
    \item \makebox[3.5cm]{*\psrB{}  \hfill} (TCG \& Spike)       
    \item \makebox[3.5cm]{\psrC{}   \hfill} (TCG \& RMS)         
    \item \makebox[3.5cm]{*\PSRpos{}\hfill} (TCG \& RMS \& Spike)  
    \item \makebox[3.5cm]{\psrE{}   \hfill} (TCG \& RMS)         
    \item \makebox[3.5cm]{\psrF{}   \hfill} (TCG \& RMS)         
\end{enumerate}


The periods of pulsars are much shorter than our time resolution --- in fact, \psrC{} and \psrG{} are millisecond pulsars \citep{johnston1993discovery, keith2011discovery} --- so it was either minute-timescale brightness fluctuation that was detected, unresolved giant pulses, or simply the scintillation of their time-averaged brightness \citep{10.1093/mnras/stw1871}, rather than individual pulses. At least \psrF{} is known to null on timescales of minutes \citep{huguenin1970radio}, and well studied with the MWA \citep[e.g.][]{mcsweeney2017phase}, so it is possible that its variability due to nulling was what we detected. The starred items are discussed in subsequent subsections.

Of the candidates which fell below the cutoff in \autoref{fig:group_size_vs_group_sn}, three of their coordinates coincided with known pulsars, but after visual inspection they were indistinguishable from the overwhelming majority of dim candidates. It is difficult to determine whether these are chance coincidences, but since they do not pass visual inspection they do not represent false negatives. They would have had to be detected at least 5 to 15 times rather than only once to be convincing given the criteria in \autoref{subsec:grouping} and how faint they are.

\subsection{GLEAM-X J0704\ensuremath{-}37}\label{subsec:gleamxj0704-37} 

\begin{figure*}[hbt!]
    \centering
    \includegraphics[width=0.71\linewidth]{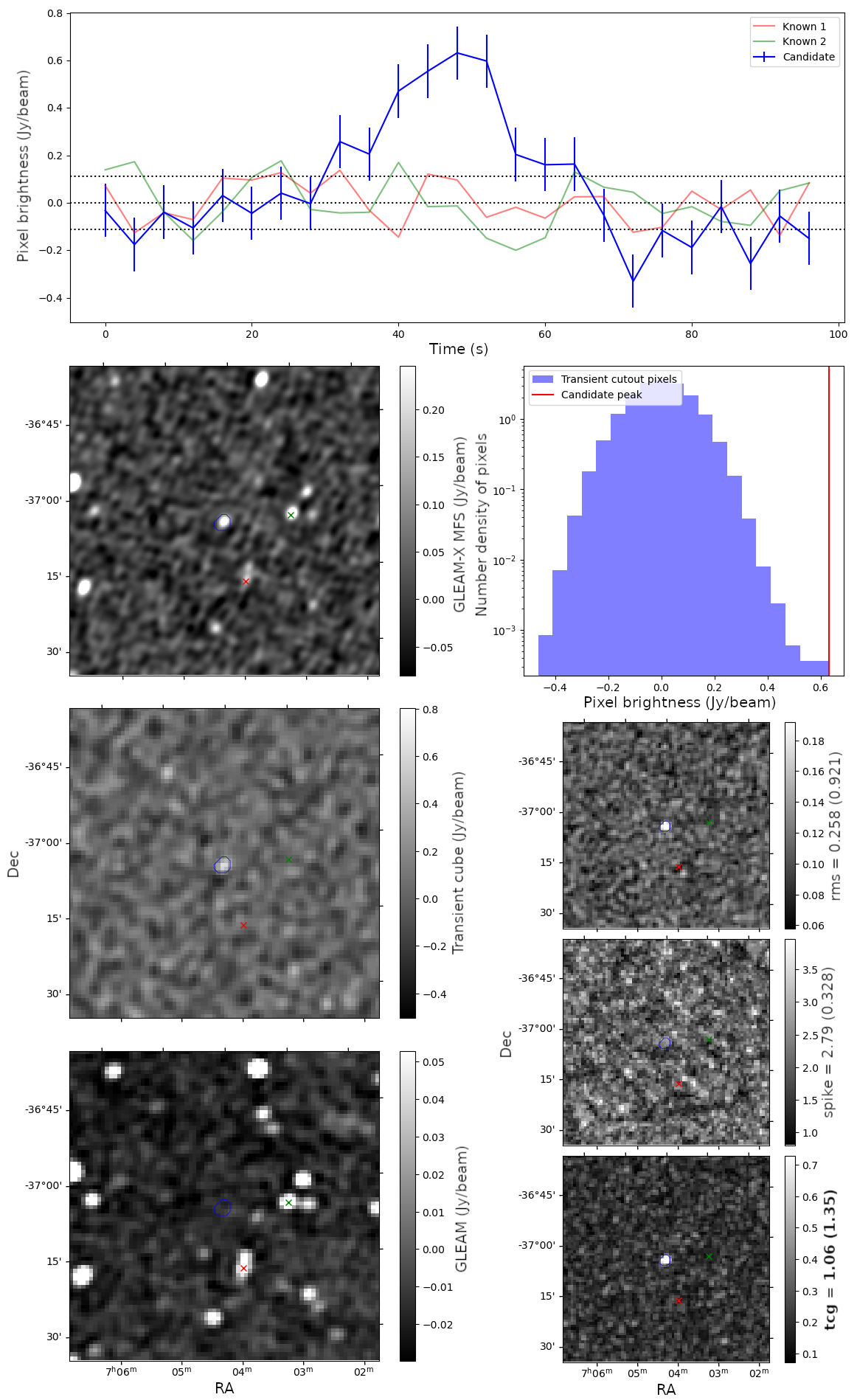}
    \caption{Diagnostic plot for the detection of \GLXLPT{}. The top panel is the lightcurve of the detected island (blue), and the lightcurves of the two nearest known sources (red and green). The dashed horizontal lines are the observation mean pixel brightness (centre line) and the positive and negative RMS. In the left column are 1\textdegree{} cutouts from the multi-frequency synthesis (MFS) image formed during the routine GLEAM-X imaging (top), peak time-step from the model-subtracted data cube (middle), and GLEAM image cutout (bottom). The blue contour marks the candidate island, and the crosses are the nearest known sources. At the top of the right column is a histogram of the pixel values in the model-subtracted cube with the candidate peak marked in red. Below are 1\textdegree{} cutouts from the RMS, Spike, and TCG filter maps, with the filter that triggered the detection in bold (in this case TCG). The numbers following the filter name in the vertical labels are the peak filter value followed by the filter value divided by the upper threshold from \autoref{tab:filter_settings} in parentheses.}
    \label{fig:glxlpt}
\end{figure*}

\GLXLPT{} was discovered using an earlier iteration of the method described here in an archival observation centered on \qty{118}{\mega\hertz} of GLEAM-X DRI and DRII. \autoref{fig:glxlpt} is the diagnostic candidate plot for the detection of \GLXLPT{}, which was at the very edge of the searched area (see \autoref{fig:sky_map_cands}).

In follow-up observations and archival observations published by \citet{Hurleywalker2024}, it was found to be a long-period radio transient, with a periodicity of $P=\qty{2.9}{\hour}$ and a DM of 36.54(1)\,\DMunit{}.
An optical counterpart was found, consistent with a cool main-sequence star of spectral type M3; further follow-up by \citet{rodrigues2025} identified a white dwarf companion, characterising this system as a polar M-dwarf / WD in a tight, tidally-locked $P=\qty{2.9}{\hour}$ orbit.

\subsection{PSR J0410\ensuremath{-}31}\label{subsec:psrj0410-31}

\PSRpos{} was discovered in a single-pulse search of the High Time Resolution Universe (HTRU) survey with the Parkes ``Murriyang'' radio telescope at 1.352\,GHz \citep{2011MNRAS.416.2465B}. It was found to have a periodicity $P=1.8785(2)$\,s and a dispersion measure (DM) of 9.2(3)\,\DMunit{}. Confirmed by \citet{2016MNRAS.455.2207R}, also with Murriyang, this source is an intermittent Rotating Radio Transient (RRAT), capable of producing high-fluence bursts and sometimes invisible (presumably nulling) in standard periodicity searches.

\PSRpos{} was detected in our analysis as a series of $\sim$Jy-bright unresolved pulses, consistent with its RRAT classification. It was observed to lie $\sim$10$'$ away from its catalogued location, likely due to the large positional error of Parkes, and the low accuracy of timing solutions generated for RRATs. To further improve the localisation, we selected the ObsID 1201778784, the highest-frequency (170--200\,MHz) observation with the brightest (1.8-Jy; S/N$\sim$45) pulse. We imaged the two 0.5-s timesteps in which the brightest pulse was detected, performed source-finding with \textsc{Aegean} \citep{agean} and compared the nearby radio galaxy positions to the Sydney University Molonglo Sky Survey \citep[SUMSS;][]{2003MNRAS.342.1117M} and the NRAO VLA Sky Survey \citep[NVSS;][]{1998AJ....115.1693C}. For the 30 S/N$>10$ sources in a $4^\circ\times4^\circ$ region around the pulsar, bulk shifts were observed, likely due to ionospheric refraction \citep{hurley2018distorting}. Against SUMSS, $\Delta$RA$=-13\farcs8\pm7\farcs2$, and $\Delta$Dec$=-0\farcs7\pm3\farcs5$; against NVSS, $\Delta$RA$=-14\farcs7\pm7\farcs6$, and $\Delta$Dec$=-1\farcs8\pm4\farcs3$. We therefore applied corrections of 14$''$ and $1''$ to our detected position to account for these shifts, and expect a residual astrometric error of $\sim3''$. Our updated position of \PSRpos{} is therefore 04$^\mathrm{h}$09$^\mathrm{m}$55.5$^\mathrm{s}$ $-31^\circ10'40\farcs5$, which is consistent with the original position reported by \citet{2011MNRAS.416.2465B}. Use of these updated coordinates should slightly improve the S/N of future pulsation and single-pulse searches towards this source.

\subsection{PSR J0031\ensuremath{-}57}\label{subsec:psrj0031-37}

Similar to \PSRpos{}, \psrB{} was also detected as a series of unresolved bursts with a small but discernible DM of ${\sim}6\,$\DMunit{}, consistent with a Galactic pulsar or RRAT. However, no source at that location was found in existing pulsar or RRAT catalogues. Archival MWA data were then searched, yielding detections of \psrB{} in high-time resolution MWA data in both 2018 and 2020. These observations confirmed that \psrB{} is a pulsar with spin period $P = 1.57\,$s and ${\rm DM} = 6.8\,$\DMunit{}. Despite being discovered via its relatively rare, unusually bright pulses, \psrB{} also exhibits much dimmer pulses that appear to be persistent, casting doubt on whether it should be classified as an RRAT. A full description of these results is given by \citet{Mcsweeney_2025}.

\section{Evaluation of Performance}\label{sec:evaluation_of_performance}

\begin{figure}[hbt!]
    \centering
    \includegraphics[width=\linewidth]{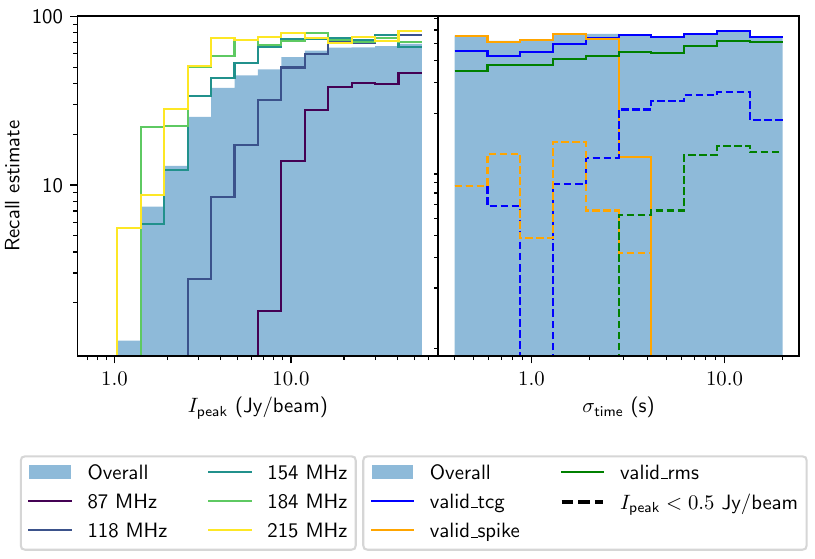}
    \caption{Histograms of injected modelled transients which were recovered after rejecting flagged candidates, as a percentage of the total number injected to estimate the recall. The injected pulse profiles have a Gaussian shape with standard deviation $\sigma_\text{time}$ and peak pixel brightness $I_\text{peak}$.}
    \label{fig:mod_sel_det_hist}
\end{figure}

The filter parameters in \autoref{tab:filter_settings} were optimised to maximise detection probability while minimising the false positive rate in a sample of 100 observations from the dataset, 20 from each frequency band, which did not contain any genuine transients. The sample was searched to estimate the false positive rate, the results of which are presented in \autoref{subsec:fale_positives}. Into a copy of the same 100 observations, modelled transients were injected. The copy was searched to estimate the true positive rate, the results of which are summarised in \autoref{subsec:modelled_transients}.

Standard statistics commonly used in the context of automatic classifiers such as precision, recall, and accuracy are difficult to use in this context. The recall estimate (fraction of real transients detected) depends on the distribution of injected transients in terms of brightness and duration, as well as the density of other sources on a particular region of the sky and the wavelengths observed. 
\autoref{fig:mod_sel_det_hist} attempts to disambiguate these dependencies, except for the spatial one. We see that at 87\,MHz the curve of recall against peak pixel brightness begins to plateau below 50\% which, combined with the large number of candidates it generates, was the justification for its exclusion by the invalid\_freq flag.

Regarding TCG, the choice of $\sigma_\text{tcg} = 4$ s corresponds to the kernel representing a pulse template $\sim 6 \sigma_\text{tcg} = 24$ s wide, about a quarter of the observation duration. In \autoref{fig:mod_sel_det_hist} we see that the TCG recall biases longer pulses as desired, and is successful even for injected transients with pulse widths $\sigma_\text{time} \leq 5 \times \sigma_\text{tcg}$. This kernel width was found to be a good compromise to maximise sensitivity between the shorter Spike sensitivity timescale and the longer RMS sensitivity timescale (although RMS was not initially conceived for this purpose).

The precision (fraction of candidates which are real transients) and accuracy (fraction of correctly classified samples) lose their meaning entirely because a) we do not have a good idea of the density of real transient sources relative to artefacts (the density of injected transients is arbitrary), b) we expect a very low precision in the first place for a blind search like this (and this is not a problem), and c) the distinction between classified samples is ambiguous because our method treats the sky as a continuous area.

\subsection{False Positives}\label{subsec:fale_positives}

The transient search steps were applied to the 100 sample observations. \autoref{fig:real_per_type_mat} is a confusion matrix of the \autoref{subsec:candidate_selection} flags assigned to the candidate islands found in the sample observations, and \autoref{fig:real_sel_per_type_mat} is a confusion matrix after all flagged candidates have been removed. All of these candidates are assumed not to be true transients, and the down-selected candidate set was visually inspected to verify this.

There were 6693, 404, and 5485 islands detected by TCG, Spike, and RMS respectively, of which 41, 199, and 6 made it through the flagging. There was effectively no overlap between islands detected by Spike and the other filters owing to the much shorter timescale of sensitivity, while TCG and RMS overlapped by over 80\%. While more than 10 times as many islands were detected by TCG and RMS than Spike, Spike had the most candidates after flagging. Spike candidates are often the result of random fluctuations in the background thermal noise, but even a low S/N detection could be real if it reoccurs multiple times. For this reason, rather than raising the Spike cutoff threshold, we allowed the grouping step in \autoref{subsec:grouping} to reject the majority of low S/N spike detections.

The greatest number of islands were rejected by the scintil\_dist flag for being too close to a known source, and most of those were from the TCG or RMS filters. Unfortunately, the TCG filter is especially sensitive to variations at exactly the timescale of ionospheric scintillation. These effects are the worst at the lowest frequencies due to the square relationship with wavelength, which is the reason for scaling the cutoff at low frequencies for scintil\_dist in \autoref{subsec:candidate_selection}. This relationship between false positive rate and frequency is visible in the gradient across the frequency columns in the top three rows (or lack thereof for Spike) of \autoref{fig:real_per_type_mat}. The Spike false positive rate does not correlate with observing frequency because it does not tend to pick up scintillation due to its short timescale sensitivity.
The sensitivity of RMS and TCG to scintillation is evident in the large values at the intersection of rows scintil\_dist and scintil\_corr, and columns valid\_tcg and valid\_rms in the confusion matrix. Their sensitivity to the long synthesised beam sidelobes around sources is evident in their intersection with invalid\_majmin.

\subsection{Modelled Transients}\label{subsec:modelled_transients}

Transients were modelled as a Gaussian pulse through time defined by the peak pixel brightness $I_\text{peak}$ and the pulse profile standard deviation $\sigma_\text{time}$, multiplied by a spatial 2D circular Gaussian with $\sigma_\text{space} = 1 \text{pixel}$. When adding a transient to a data cube, the mean through time was subtracted to simulate the model subtraction step. For $N_\text{inj}$ transients at time $t_i$ and pixel coordinates $(x_i,y_i)$ injected into an observation $I_\text{obs}$, the resulting data cube $I_\text{inj}$ is described by:
\begin{equation}\label{eqn:modelled_transient}\begin{aligned}
    &I_\text{trans}^i(t,x,y) =\\&I_\text{peak} \text{exp}\left(-\frac{(t-t_i)^2}{2\sigma_\text{time}^2} - \frac{(x-x_i)^2 + (y-y_i)^2}{2\sigma_\text{space}^2} \right)
\end{aligned}\end{equation}
\begin{equation}\label{eqn:add_modelled}\begin{aligned}
    &I_\text{inj}(t,x,y) =\\&I_\text{obs}(t,x,y) + \sum_{i=1}^{N_\text{inj}} \left( I_\text{trans}^i(t,x,y) - \left< I_\text{trans}^i(t',x,y) \right>_{t'} \right)
\end{aligned}\end{equation}

A circular Gaussian is a good approximation because we do not expect transients bright enough to produce significant sidelobes with the MWA's $(u,v)$-coverage, and $\sigma_\text{space} = 1 \text{pixel}$ corresponds to a FWHM of $2.355$\,pixels which is approximately the PSF FWHM. Although an even distribution of modelled source brightnesses in an image is not realistic, $I_\text{peak}$ was not multiplied by the primary beam because statistics relative to the background noise level (which does not vary with the primary beam) are more useful in this analysis.

In each observation, $N_\text{inj} = 100$ modelled transients were injected with uniform distributions of position, $I_\text{peak} \in [0.1, 3.0]$ Jy, and $\sigma_\text{time} \in [0.4, 20]$ s. Note that $\sigma_\text{time} = 20$ s corresponds to a pulse width of $\sim 120$ s if we consider the a Gaussian pulse to extend to $\sim 3 \sigma_\text{time}$, which is the observation length. Pulses on a longer timescale would not be imaged anyway due to the model subtraction (\autoref{subsec:imaging}). The same filtering and flagging steps as before were applied to the data cubes with injected transients.

\begin{figure*}[ht!]
\centering
    \begin{subfigure}{.545\textwidth}
        \centering
        \includegraphics[width=\linewidth,trim={1.1cm 0 1.9cm 0},clip]{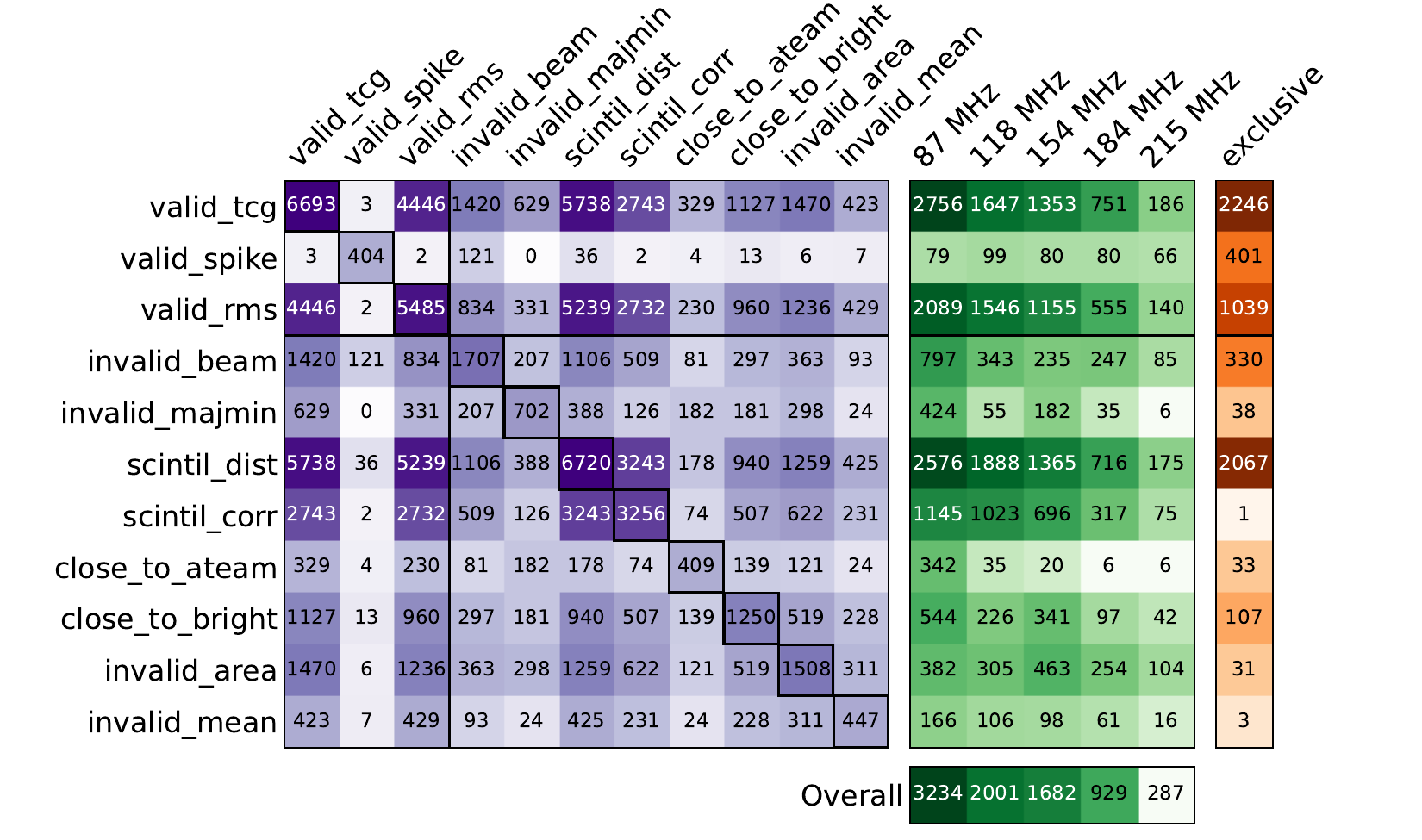}
        \caption{Candidate islands detected in the observation sample. All candidates bellow row 3 \newline{}(horizontal black line) are to be rejected.}
        \label{fig:real_per_type_mat}
    \end{subfigure}
    \hfill
    \begin{subfigure}{.445\textwidth}
        \centering
        \includegraphics[width=\linewidth,trim={5.6cm 0 1.9cm -2cm},clip]{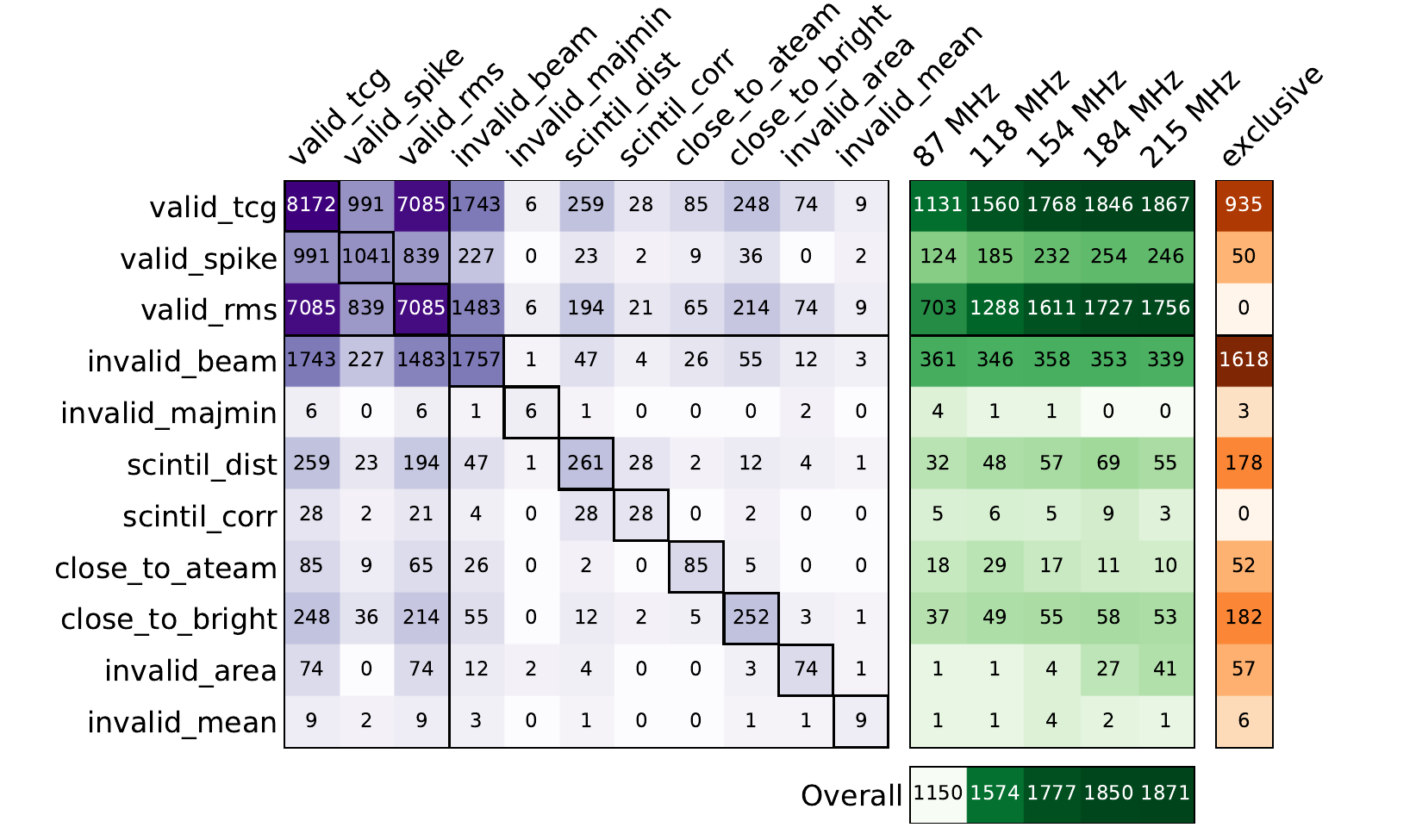}
        \caption{Injected modelled transients which were recovered. All candidates bellow row 3 (horizontal black line) were incorrectly flagged as artefacts.}
        \label{fig:mod_per_type_mat}
    \end{subfigure}

    \vspace{1em}
    
    \begin{subfigure}{.45\textwidth}
        \centering
        \includegraphics[width=0.9\linewidth]{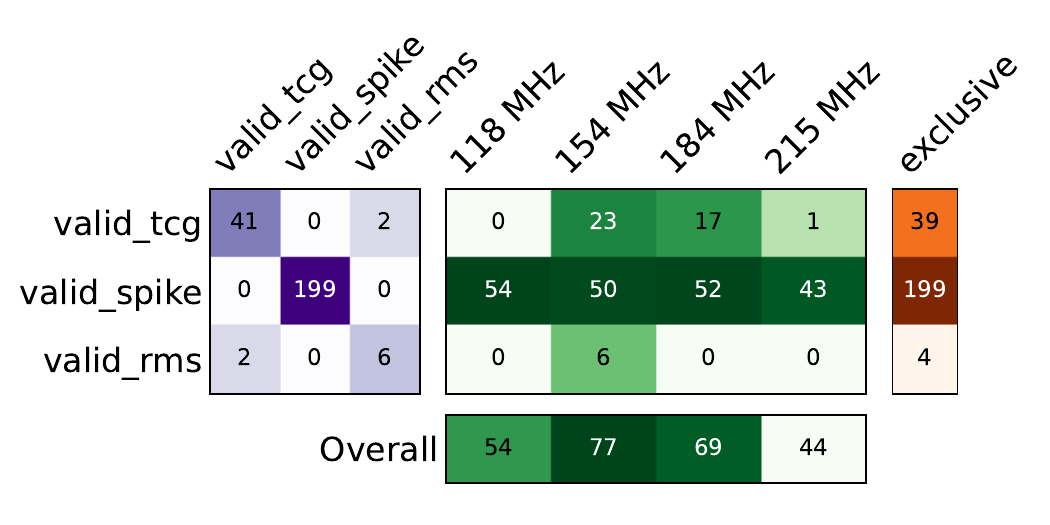}
        \caption{Candidate islands detected in the observation sample after rejecting flagged candidates.}
        \label{fig:real_sel_per_type_mat}
    \end{subfigure}
    \hfill
    \begin{subfigure}{.45\textwidth}
        \centering
        \includegraphics[width=0.9\linewidth]{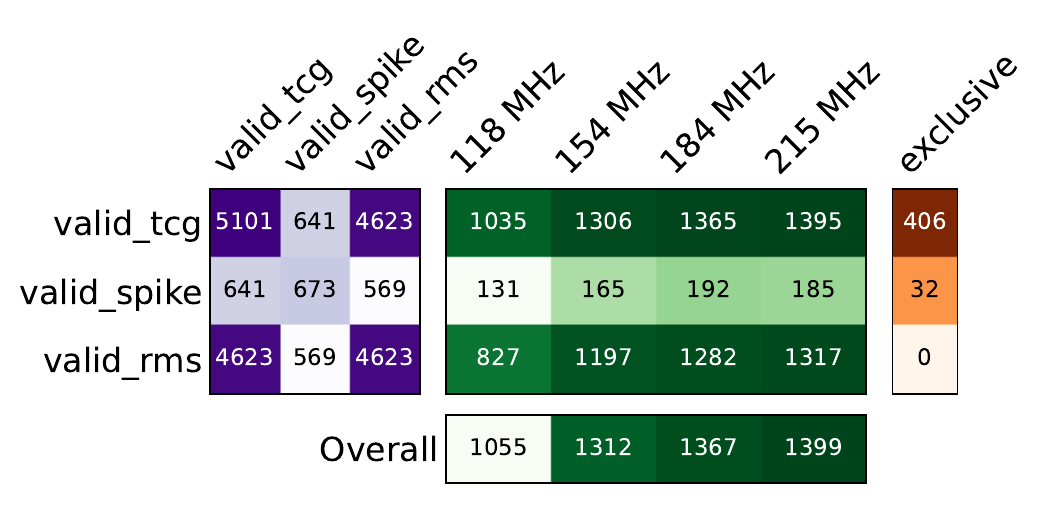}
        \caption{Injected modelled transients which were recovered after rejecting flagged candidates.}
        \label{fig:mod_sel_per_type_mat}
    \end{subfigure}

    \caption{Confusion matrices of candidate flags detected in a sample of 100 observations (20 at each of the 5 frequencies). Each cell counts the number of candidates in both the row category AND the column category. The left square matrices count the flags assigned to candidates, and the values along the diagonal are the total with that flag. The middle matrices count the flags in observations in different frequency channels, with the bottom ``Overall'' row counting the total number of candidates in each channel. The ``Exclusive'' column at right counts the number of candidates with only a particular flag (i.e. 38 candidates were rejected purely because of invalid\_majmin in (a)). Note that is\_moon wasn't included because the moon wasn't in this subset of observations, and invalid\_freq wasn't included because it is equivalent to the 87\,MHz column.}
    
\end{figure*}

The number of islands detected which corresponded with injected transients with each flag are shown in \autoref{fig:mod_per_type_mat} in the same format as \autoref{fig:real_per_type_mat}, and the number recovered after rejecting flagged islands is shown in \autoref{fig:mod_sel_per_type_mat}. The majority of false negatives were due to the invalid\_beam flag. This is unavoidable because the edges of the observations where sensitivity is low are too noisy to trust detections. The next most common were scintil\_dist and close\_to\_bright due to injected transients falling on real sources by chance. After candidate selection, 64\% of modelled transients were recovered. For the 118\,MHz band, only 53\% were recovered, while for 154\,MHz and above, 69\% were recovered. TCG, Spike, and RMS detected 64\%, 8.4\%, and 59\% of injected transients respectively, but RMS detected only 1 which was not also detected by either Spike or TCG. The right panel of \autoref{fig:mod_sel_det_hist} shows that spike is most sensitive on short timescales, while TCG and RMS are sensitive on long timescales, especially when it comes to dim sources.

\section{Conclusion}

In this work we have developed and tested a new method for searching for radio transients on the minute timescale. Testing the method on the GLEAM-X DRI and DRII we discovered a new pulsar \psrB{} and a new LPT \GLXLPT{} which, due to its high Galactic latitude out of the busy Galactic plane, could for the first time be associated with an optical counterpart.
Due to the high latitudes of most of the searched data (see \autoref{fig:sky_map_obs}) we might expect a relatively low number density of real transients, making these discoveries all the more interesting.
Nevertheless, it served as a diverse and extensive dataset of false positives to develop and tune this algorithm. The MWA Galactic plane monitoring programme is a more targeted search, with observations at 185–215\,MHz covering $|b| < 15^\circ$ and $73^\circ < l < 284^\circ$ (Hurley-Walker et al. in prep) in 2022 (with a second phase in 2024/25). It discovered the $P\sim$22-min LPT GPM\,J1839\ensuremath{-}10 \citep{Hurleywalker2023} and confirmed ASKAP\,J1755\ensuremath{-}25 \citep{2025MNRAS.tmp.1156M} as a $P\sim$1.16 hour LPT using iterations of the method described in this paper.

The number of hand-tuned parameters and decision layers in this search method prompts the consideration of machine learning (ML). While beyond the scope of this work, an ML model --- perhaps image-based, decision-tree-based, or otherwise --- could be trained on a dataset consisting of synthetic transients and real artefacts generated by our method. A pre-selection of candidates would still be necessary in such a method because convolving an ML model over the entire sky area would be prohibitively expensive.

The most effective methods for separating the true positives from the false positives were the scintil\_dist flag (see the ``exclusive'' column of \autoref{fig:real_per_type_mat} and \autoref{fig:mod_per_type_mat}) and the candidate grouping (see \autoref{fig:group_size_vs_group_sn}). The latter highlights that the best way to be sure a repeating source is real is to see it multiple times, but whether or not a given source is detected multiple times is a complicated function of observation length and observing cadence. Generally, however, the chances of re-detection go down for longer period sources (see \ref{sec:period_transient_detectability}). The small number of candidates with multiple detections did not justify the implementation of any automated periodicity detection in this study, but this may be effective in future larger surveys.

\begin{acknowledgement}
This scientific work uses data obtained from Inyarrimanha Ilgari Bundara, the CSIRO Murchison Radio-astronomy Observatory. Support for the operation of the MWA is provided by the Australian Government (NCRIS), under a contract to Curtin University administered by Astronomy Australia Limited. ASVO has received funding from the Australian Commonwealth Government through the National eResearch Collaboration Tools and Resources (NeCTAR) Project, the Australian National Data Service (ANDS), and the National Collaborative Research Infrastructure Strategy.

This work was supported by resources provided by the Pawsey Supercomputing Centre, which is also used by the MWA. Establishment of Inyarrimanha Ilgari Bundara, and the Pawsey Supercomputing Centre are initiatives of the Australian Government, with support from the Government of Western Australia and the Science and Industry Endowment Fund. We acknowledge the Wajarri Yamaji People as the Traditional Owners and Native Title Holders of the observatory site.

We acknowledge ADACS and Dr Paul Hancock for the development of the transient classification web-interface.
\end{acknowledgement}

\paragraph{Funding Statement}
This research is supported by an Australian Government Research Training Program (RTP) Scholarship {doi.org/10.82133/C42F-K220}.
N.H.-W. is the recipient of an Australian Research Council Future Fellowship (project number FT190100231).

\paragraph{Competing Interests}
The authors declare no competing interests.

\paragraph{Data Availability Statement}
A FITS table of candidates will be made available on the PASA Datastore. Further data used in this work is available upon request from the authors.

\paragraph{Code Availability Statement}
The code used in this work is available at \url{https://github.com/CsanadHorvath/Transients.git}.

\printendnotes

\bibliography{example}

\appendix

\newpage
\section{Periodic Transient Detectability}\label{sec:period_transient_detectability}

\begin{figure}[hbt!]
    \centering
    \includegraphics[width=\linewidth]{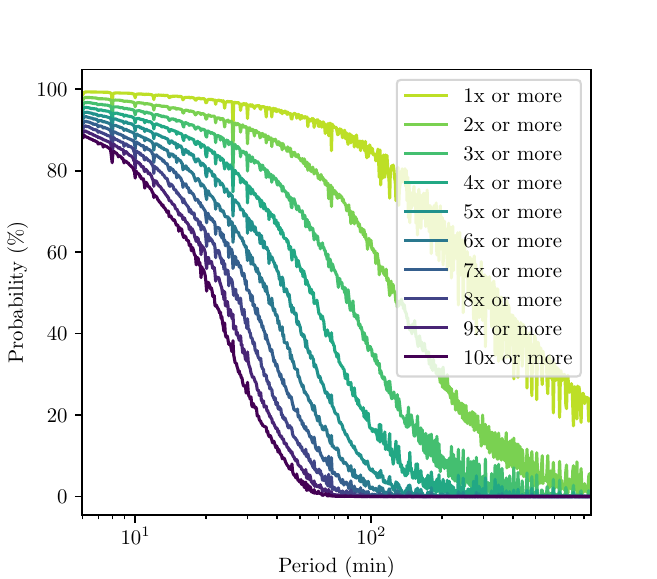}
    \caption{Probability that a periodic transient is pointed towards Earth and in the MWA field x number of times in GLEAM-X  DRI and DRII.}
    \label{fig:coverage_prop}
\end{figure}

To get a sense of the likelihood of a periodic source being in the field of view while it is active at all, we ran a simple simulation as follows. For an even distribution of coordinates in the dataset coverage shown in \autoref{fig:sky_map_obs} we counted the number times a periodic source at that coordinate would have both been on and in the field. In \autoref{fig:coverage_prop} we show the fraction of those hypothetical sources which would have been seen at least some number of times. We see that a $\sim$10 min source has a $\sim$95\% chance of being in the field at least 5 times, whereas a $\sim$200 min source like \GLXLPT{} has a $\sim$60\% chance of being seen just once. The vertically aligned spikes in the curves are caused by harmonics between the periods and observing times. Critically, this ignores the actual detectability of sources; LPTs are difficult to find even with a perfect telescope. The problem is further confounded by the fact that many LPTs have very short (sometimes repeating) windows of activity.

\vspace{-5px}

\section{Extended Figures}

\begin{figure*}[hbt!]
    \centering
    \includegraphics[width=0.85\linewidth]{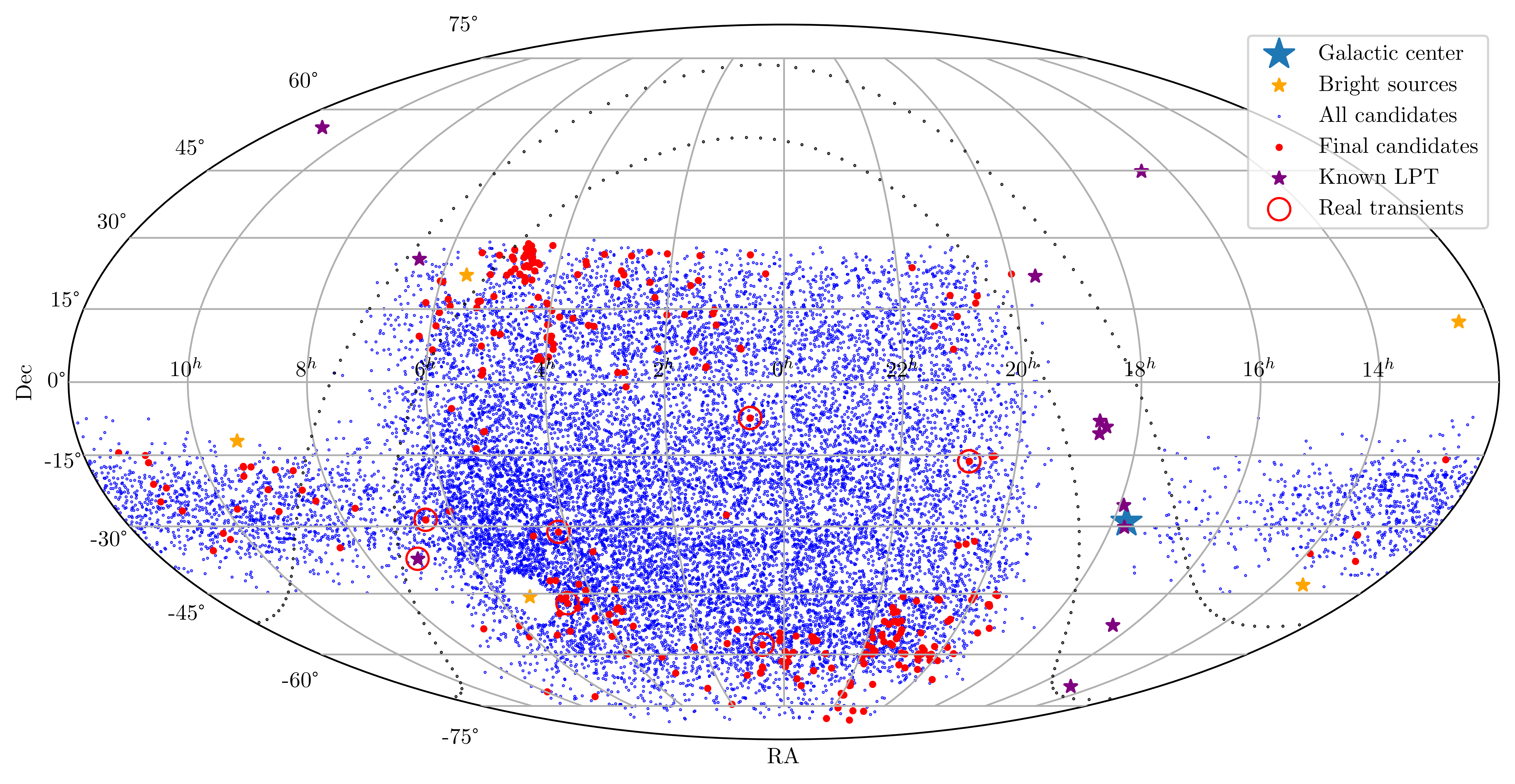}
    \caption{Sky map of transient candidates from GLEAM-X  DRI and DRII. The blue points are the candidate groups after artefact flagging in {\autoref{subsec:candidate_selection}. The red points are the final candidates after applying candidate grouping in {\autoref{subsec:grouping}}. The bright sources are those listed under close\_to\_ateam in {\autoref{subsec:candidate_selection}}.}
}
    \label{fig:sky_map_cands}
\end{figure*}

\begin{figure*}[hbt!]
    \centering
    \includegraphics[width=0.95\linewidth]{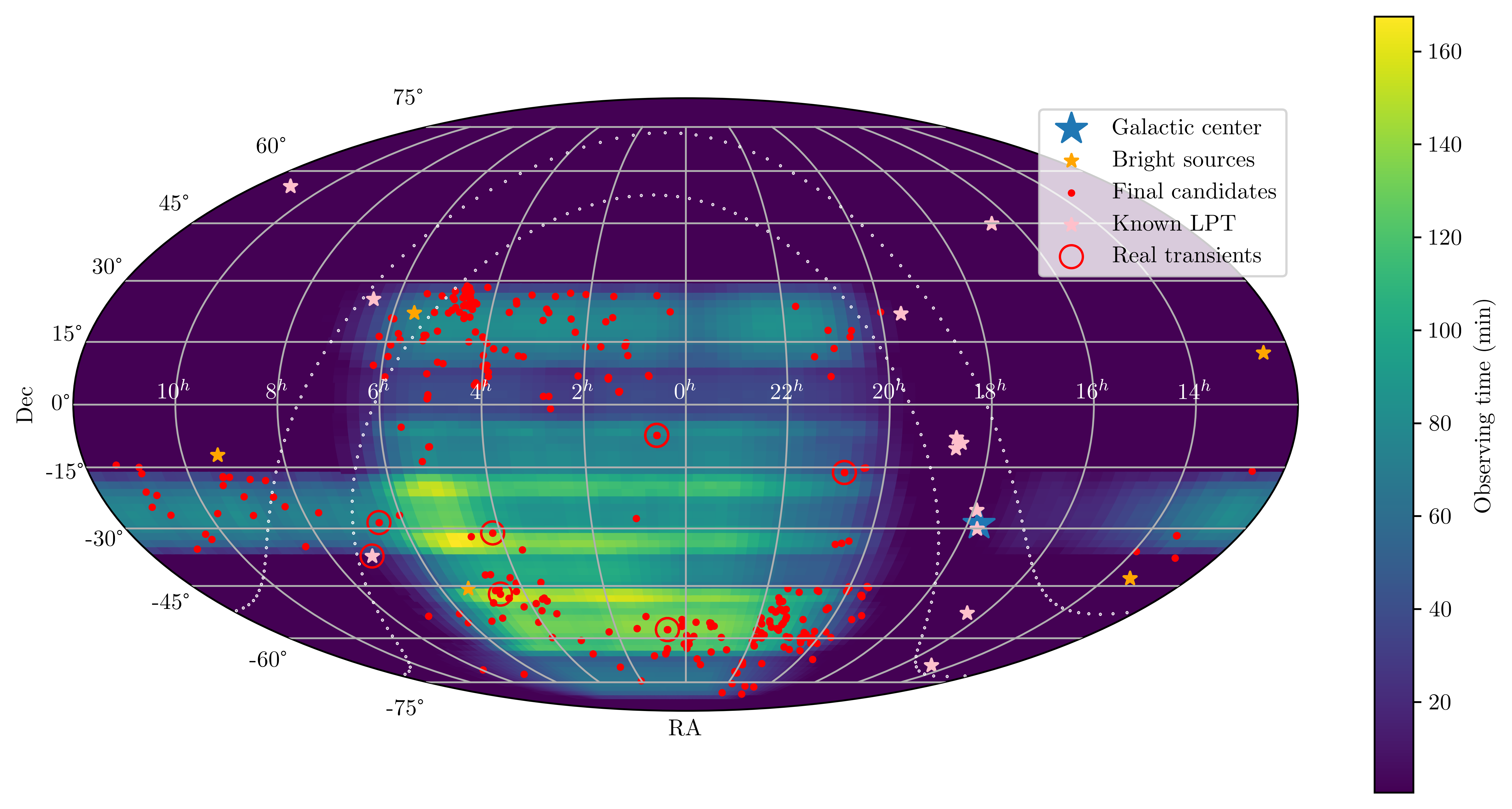}
    \caption{Sky map of GLEAM-X DRI and DRII observations. The red points are the final candidates after applying candidate grouping in {\autoref{subsec:grouping}}. The colour map is the total (non-continuous) minutes of data at a coordinate. The bright sources are those listed under close\_to\_ateam in {\autoref{subsec:candidate_selection}}.}
    \label{fig:sky_map_obs}
\end{figure*}

The LPTs shown in {\autoref{fig:sky_map_cands}} and {\autoref{fig:sky_map_obs}} are:
\begin{itemize}
    \item CHIME      \,J0630{\ensuremath{+}}25 {\citep{dong2024discovery}}
    \item GLEAM-X    \,J0704{\ensuremath{-}}37 {\citep{Hurleywalker2024}}
    \item ILT        \,J1101{\ensuremath{+}}55 {\citep{de2025sporadic}}
    \item ASKAP      \,J1448{\ensuremath{-}}68 {\citep{2025MNRAS.tmp.1214A}}
    \item GLEAM-X    \,J1627{\ensuremath{-}}52 {\citep{Hurleywalker2022}}
    \item CHIME/ILT  \,J1634{\ensuremath{+}}44 {\citep{2025ApJ...988L..29D,2025A&A...699A.341B}}
    \item GCRT       \,J1745{\ensuremath{-}}30 {\citep{2005Natur.434...50H}}
    \item ASKAP      \,J1755{\ensuremath{-}}25 {\citep{2024MNRAS.535..909D,2025MNRAS.tmp.1156M}}
    \item ASKAP/DART \,J1832{\ensuremath{-}}09 {\citep{2024arXiv241115739L,wang2025detection}}
    \item GPM        \,J1839{\ensuremath{-}}10 {\citep{Hurleywalker2023}}
    \item ASKAP      \,J1839{\ensuremath{-}}07 {\citep{lee2025emission}}
    \item ASKAP      \,J1935{\ensuremath{+}}21 {\citep{caleb2024emission}}
\end{itemize}

\begin{table*}[hbt!]
    \caption{
        Properties of the model-subtracted image cubes towards zenith for each frequency band. All images are $2400 \times 2400$\,pixels.
    }
    \label{tab:data_properties}
    \begin{tabular}{llllll}
        \toprule
        \headrow                      & 88\,MHz        & 118\,MHz      & 154\,MHz      & 185\,MHz      & 216\,MHz \\
        \midrule
             PSF FWHM semi-major axis & 156$''$       & 115$''$      & 87$''$       & 72$''$       & 62$''$      \\
                                      & 2.5\,pix      & 2.5\,pix     & 2.5\,pix     & 2.5\,pix     & 2.5\,pix     \\
             PSF FWHM semi-minor axis & 121$''$       & 89$''$       & 68$''$       & 56$''$       & 48$''$      \\
                                      & 1.9\,pix      & 1.9\,pix     & 1.9\,pix     & 1.9\,pix     & 1.9\,pix     \\
             Pixel size               & 63$''$        & 47$''$       & 36$''$       & 30$''$       & 26$''$      \\
             Image sky area           & 1742\,deg$^2$ & 959\,deg$^2$ & 567\,deg$^2$ & 394\,deg$^2$ & 290\,deg$^2$     \\
        \bottomrule
    \end{tabular}
\end{table*}

\end{document}